# The Role of Mutation Rate Variation and Genetic Diversity in the Architecture of Human Disease

Ying Chen Eyre-Walker
Adam Eyre-Walker*

School of Life Sciences
University of Sussex
Brighton
BN1 9QG

*Correspondence: a.c.eyre-walker@sussex.ac.uk

**Abstract**

*Background*

We have investigated the role that the mutation rate and the structure of genetic variation at a locus play in determining whether a gene is involved in disease. We predict that the mutation rate and its genetic diversity should be higher in genes associated with disease, unless all genes that could cause disease have already been identified.

*Results*

Consistent with our predictions we find that genes associated with Mendelian and complex disease are substantially longer than non-disease genes. However, we find that both Mendelian and complex disease genes are found in regions of the genome with relatively low mutation rates, as inferred from intron divergence between humans and chimpanzees. Complex disease gene are predicted to have higher rates of non-synonymous mutation than non-disease genes, but the opposite pattern is found in Mendelian disease genes. Finally, we find that disease genes are in regions of significantly elevated genetic diversity, even when variation in the rate of mutation is controlled for. The effect is small nevertheless.

*Conclusions*

Our results suggest that variation in the genic mutation rate and the genetic architecture of the locus play a minor role in determining whether a gene is associated with disease.

**Introduction**

Why do humans suffer from the diseases that we do? In part this is clearly due to our anatomy and physiology, and that of the organisms that infect us - we cannot have a disease of an organ that we do not possess. But why do we suffer from cystic fibrosis rather than some other disease of the lungs? One simple reason might be variation in the mutation rate. Those genes and genomic regions that have high mutation rates are more likely to generate disease mutations, and hence be associated with a disease. The rate of mutation of a locus will depend upon two factors: the rate of mutation per site and the number of sites at which a mutation can generate a disease phenotype. The per site mutation rate is known to vary across the human genome at a number of different scales such that some genes have mutation rates that are several fold higher than other genes (reviewed in [1]). Genes also vary considerably in their length, with some of the largest, such as the dystrophin gene, being association with disease.

A more subtle factor affecting the likelihood of a gene being associated with a disease is the genealogy. At each site in the genome there is an underlying genealogy whereby every chromosome in the population is related via a bifurcating tree to every other chromosome at that site. If there is no recombination between sites then sites share the same genealogy. The shape and depth of the genealogy depends on several factors. The first is chance; for example, the average total length of a genealogy for a neutral locus in a population of stationary size is expected to be proportional to $4N$ generations in a diploid species, where $N$ is the population size, but this is expected to have a variance of at least $(4N)^2$ generations [2]. Second, the genealogy depends on the effective population size of the locus ($N_e$). $N_e$ is thought to vary across the human genome as a consequence of natural selection [3,4]. Selection can reduce the $N_e$ of a genomic region through either a selective sweep caused by the passage of an advantageous mutation through the population [5], or via background selection caused by the removal of deleterious mutations [6]. Those regions of the genome with low rates of recombination or a high density of selected sites are expected to have low $N_e$, and this is expected to reduce the genetic diversity of neutral and weakly selected variants in these regions (reviewed in [7]). Analyses suggest that $N_e$ varies across the human genome by a few-fold [4]. The effective population size is not expected to affect the frequency of deleterious mutations in which the product of $N_e$ and the strength selection is greater than one. However, stochastic factors are expected to be important irrespective of the selection acting upon a mutation.

Previous analyses have shown that Mendelian disease genes are 30% longer than non-disease genes [8,9]. Comparative analyses have also shown that genes associated with Mendelian diseases have significantly, but only slightly higher rates of mutation per site, as inferred from levels of synonymous divergence between species [8,10]. The rather modest differences between disease and non-diseases genes in the inferred mutation rate might be due to time frame over which the mutation rate was inferred: Smith and Eyre-Walker [8] considered the divergence between human and mouse, and Huang et al. [10] considered the divergence between mouse and rat. This will give a poor estimate of the current mutation rate at a locus in humans because the relative mutation rate of a locus appears to have evolved through time [1,11]. The mutation rate has also recently been predicted, based on a model fitted to the locations of *de novo* mutations in humans, to be slightly higher in disease associated genes [12], but the accuracy of this model is unproven, and they consider the total mutation rate of the exon, rather than the rate at non-synonymous sites. Here we consider the divergence between humans and their most closely related extant relative, chimpanzee, as our measure of the mutation rate. We also consider whether the density of single nucleotide polymorphism is greater in disease than non-disease genes.

**Results**

We predict that unless all possible diseases with a genetic basis, and all the genes that can cause them, have already been discovered, then genes associated with diseases should have genic mutation rates than non-disease genes, where the genic mutation rate is determined by the product of gene length and the mutation rate per site. We also predict that disease genes should be in relatively diverse regions of the genome. To investigate these predictions we compiled data from 17577 autosomal genes with introns, of which 854 genes are known to cause a Mendelian disease. We also analysed 1732 genes in which the strongest signal in a genomic region in a genome wide association study (GWAS) lay within the boundaries of the gene (i.e. all exons and introns between the start and stop codon). The presence of an association signal within the boundaries of the gene does not necessarily mean that the causative mutation is within the protein coding sequence or even within the boundaries of the gene, and many of these associations may be in regulatory sequences [13]. We subsequently excluded genes on the sex chromosomes since the Y-chromosome is known to have a higher and the X-chromosome a lower mutation rate than the autosomes [14]. This yielded a dataset of 17062 genes including 820 associated with a Mendelian disease and 1726 with a GWAS signal.

*Gene length*

Consistent with the hypothesis that disease genes should have higher overall rates of mutation we find, as others have in the past for genes causing Mendelian disease [8,9], that genes associated with disease are significantly longer, in terms of coding sequence (CDS) length, than non-disease genes - disease genes are ~50% longer than non-disease genes (One-way ANOVA $p<0.001$)(Figure 1). This is greater than in previous studies [8,9], but this is likely to be due to the improvement in genome annotation. Strikingly, the difference is as great for the GWAS as Mendelian disease genes despite the fact that many of the GWAS signals are likely to be outside the protein coding sequence [13]. GWAS genes might have longer CDSs because intron and CDS lengths are correlated ($r = 0.30$, $p<0.001$), and long introns are more likely to contain a SNP that causes the disease or tags a causative mutation somewhere in the vicinity of the gene. However, if we control for the correlation between intron and CDS length by regressing CDS length against intron length and taking the residuals, we find that GWAS genes have longer CDSs, than non-disease genes, even given their longer introns (t-test p<0.001; similar results are obtained if we regress log CDS length against log intron length (p<0.001)).

*Mutation rates*

However, contrary to our expectations, we find that disease genes are found in regions of the genome with significantly lower mutation rates, as measured by intron divergence between human and chimpanzee. The difference is highly significant (one-way ANOVA p <0.001), but the difference is small with disease genes having approximately 5% lower intron divergence than non-disease genes (Figure 2a). The pattern differs between CpG and non-CpG sites, with disease genes having higher divergence at CpG sites and lower divergence at non-CpG sites (results not shown). However, if we calculate the expected non-synonymous mutation rate in the CDS by multiplying the proportion of non-synonymous sites that are CpG and non-CpG in the CDS by the respective levels of intron divergence, we find that complex disease genes have slightly higher predicted mutation rates than non-disease genes, but Mendelian disease genes have slightly lower mutation rates per site (p=0.041) (Figure 2b). As expected, both Mendelian and complex disease genes have significantly higher predicted rates of non-synonymous mutation (p<0.001), driven by the fact that disease genes have longer CDSs.

The fact that Mendelian disease genes have lower predicted rates of non-synonymous mutation per site is inconsistent with our hypothesis, but this might be due to the fact that they have features which predispose them to lower mutation rates - for example they might be transcribed at lower levels and hence have lower rates of mutation. Divergence at intronic and intergenic sites is known to be significantly correlated to a number of other variables including GC-content [3,15,16,17], recombination rate [3,16,18,19,20], replication time [15,21,22], distance to the telomere and centromere [3,15,16,23], gene density [3,16], nucleosome occupancy [11] and expression level [24]. We confirm previous results and show that intron divergence is positive correlated to GC content, replication time (later genes have higher divergence) and male recombination rate, and negatively correlated to the distance to the telomere, distance to the centromere, female recombination rate, nucleosome occupancy and germ-line expression, within a multiple regression (Table 1). Similar patterns are evident for the predicted non-synonymous mutation rate (Table 1). If we take the residuals from a multiple regression of intron divergence against all the genomic variables above we find that intron divergence does not differ significantly between disease and non-disease genes, but that the predicted rate of non-synonymous mutation is greater for both Mendelian and GWAS genes (ANOVA p<0.001), although this is only individually significant for GWAS genes (t-test p<0.001).

*Genetic diversity*

Although, disease genes are found in regions of the genome with relatively low rates of intron mutation we find that disease genes have a significantly greater density of polymorphisms segregating in their introns than non-disease genes; the difference is 11% and 17% for the Mendelian and GWAS genes respectively (Figure 3a). If we divide the density of SNPs by the divergence of introns to calculate a quantity we might call the "realized" effective population size, we find that Mendelian and GWAS genes have significantly higher realized $N_e$ values that are 9% and 12% greater than non-disease genes (ANOVA p<0.001; t-test of Mendelian versus non-disease p<0.001; t-test of GWAS versus non-disease p<0.001). However, we find no evidence that the predicted non-synonymous population mutation rate in the CDS (calculated as the proportion of non-synonymous sites that are CpG multiplied by the SNP density at CpG sites in introns plus the proportion of non-synonymous sites that are non-CpG multiplied by the SNP density at non-CpG in introns) differs between disease and non-disease genes, but the calculation of the predicted non-synonymous population mutation rate is subject to considerable error because we have relatively few intron CpG sites and SNP density is very low in humans.

As with the predicted non-synonymous mutation rate, it is possible that disease genes have higher diversities and realized $N_e$ values because disease genes have features that predispose them to higher values, not because by having higher values they are more likely to be associated with disease. We find that intron SNP density is positively correlated to GC content, female and male rates of recombination, the time of replication and distance to the centromere and negatively correlated to nucleosome occupancy, germ-line expression and distance to the telomere (Table 1). If control for these factors by taking the residuals from the multiple regression we find that SNP density is still significantly greater in both Mendelian and GWAS genes, than in non-disease genes (ANOVA p<0.001; individual t-tests p<0.001). Likewise we find the realized $N_e$ is positively correlated to all variables except GC content, nucleosome occupancy and male recombination rate (Table 1), and that after controlling for these associations, disease genes still have significantly greater realized $N_e$ values than non disease genes (ANOVA p=0.019; individual t-tests Mendelian versus non-disease p=0.21, GWAS versus non-disease p=0.001).

Although disease genes have a greater number of SNPs per bp than non-disease genes the distribution of the genetic variation varies in an inconsistent manner between categories of genes; the average minor allele frequency is ~10% greater in Mendelian, and ~10% lower in GWAS genes, than in non-disease genes (ANOVA p <0.01)(Figure 3b).

**Discussion**
We have found that genes associated with disease are longer and reside in regions of the genome with greater intron diversities and realized effective population sizes than non-disease genes. This is consistent with a role for mutation and genetic variation in determining whether a gene becomes associated with disease. However, we do not find convincing evidence that the mutation rate per site is greater in disease genes than non-disease genes; only if we consider the predicted rate of non-synonymous mutation and control for various genomic variables do we find that both Mendelian and complex disease genes have higher mutation rates than non-disease genes. Ad even then the differences seem small. Nevertheless, what is ultimately important is the mutation rate of the gene, and we find that the overall mutation rate of disease genes is greater than non-disease genes because disease genes are longer (p<0.001). The effect of gene length may be more conspicuous than for the other variables, because there is substantially more variation in CDS length per gene (coefficient of variation (CV) = 0.78) than in intron divergence (CV = 0.56), intron SNP density (CV = 0.42) and realized $N_e$ (CV = 0.47); in reality the differences in CV are even larger because intron divergence, and in particular SNP density and realized $N_e$, are likely to be subject to large sampling error variances that CDS length is not.

We have interpreted the fact that disease genes are longer than non-disease genes as evidence that genes with higher mutation rates are more likely to generate disease mutations, however, it is possible that disease genes are longer simply because genes involved in particular processes that could cause disease are longer. It is difficult to test this hypothesis without knowing all the genes that might cause disease. We have also interpreted the greater diversity in disease genes as being what causes them to be associated with disease. However, in the case of the complex disease genes this might simply reflect a bias towards a better ability to detect GWAS signals in regions of higher diversity.

Interestingly, two observations in our analysis suggest that a reasonable proportion of the causative mutations being tagged by GWAS associations may be in protein coding sequences rather than in regulatory regions. First, we find that GWAS genes have longer CDSs than non-disease genes, even if we control for the fact that genes with longer CDSs have longer total intron length. Second, we find that GWAS genes have lower intron divergence than non-disease genes, but higher predicted non-synonymous mutation rates.

Although, we have found that disease genes are longer than non-disease genes, and that they have greater diversity, realized effective population size, the differences are fairly small. It is therefore evident that either most disease associated genes have been discovered, which seems unlikely, or that the function of the gene is far more important in determining whether a gene causes disease than its effective mutation rate.

**Materials and methods**

*Genic mutation rate*

To estimate mutation rate for each gene, we estimated their intron divergence between the human and chimpanzee genomes as follows. Alignments using the NCBI build 36 version of the human genome and PanTro2 version of the chimp genome were downloaded from the UCSC website (http://genome.ucsc.edu/). Alignments were parsed into individual genic sequences and realigned with MAFFT version 6 (http://mafft.cbrc.jp/alignment/software/). Exon sequences were masked according to exon annotation of the NCBI build 36 version of the human genome from the ensemble database (http://www.ensembl.org/). We did not correct for multiple hits; this is not necessary since the average intron divergence between human and chimpanzee sequences is 1.05% [23]. We calculated the rates of intron divergence for CpG and nonCpG sites separately since the former have much higher rates of mutation. We used these intron divergences to infer the rate of non-synonymous mutation in human exons, by calculating the number of CpG and non-CpG sites in each exon which when mutated would give a non-synonymous change; in this calculation we assumed that all mutations at CpGs are transitions, which is a good aproximation [25], and that 60% of mutations at other sites were transitions. If a gene had multiple transcripts we made these calculations for each transcript and averaged the result.

*Disease genes*

Genes were designated as being associated with Mendelian disease based upon the compilation made by [26]. Genes associated with genome-wide association studies (GWAS) were obtained from GWAS catalog (http://www.genome.gov/gwastudies/); a gene in which the strongest GWAS signal was found within the boundaries of a gene were designated as being a GWAS gene.

*Genomic features*

To investigate what factors might influence patterns of genic mutations estimated by intron divergence we considered a number of variables. Intron GC content, nucleosome occupancy, replication timing and male and female recombination rates were downloaded from the UCSC website (http://genome.ucsc.edu/). We used A365 values to study the influence of nucleosome occupancy on the distribution of genic mutations rate across the genome. Recombination rates per MB were from Kong *et al* [27]. Replication time data, from HELA cells, were from Chen *et al.* [15]. Germ-line expression data were from a study by McVicker and Green [28].

**Acknowledgements**

The authors are grateful to the comments of three anonymous referees.

| Factor | Intron Divergence | Predicted non-synonymous mutation rate | Intron SNP density | Realized effective population size |
|---|---|---|---|---|
| GC content | 0.525*** | 0.512*** | 0.192*** | -0.212*** |
| Nucleosome occupancy | -0.396*** | -0.422*** | -0.142*** | -0.035 |
| Female recombination rate | -0.020* | -0.004 | 0.058*** | 0.042*** |
| Male recombination rate | 0.202*** | 0.176*** | 0.129*** | -0.048*** |
| Germ-line expression | -0.062*** | -0.090*** | -0.020* | 0.032*** |
| Replication time | 0.132*** | 0.132*** | 0.071*** | 0.038*** |
| Distance to telomere | -0.158*** | -0.154*** | -0.117*** | 0.060*** |
| Distance to centromere | -0.018* | -0.008 | 0.020* | 0.014 |

**Table 1.** Standardised regression coefficients from multiple regressions. * p<0.05, ** p<0.01 and *** p<0.001

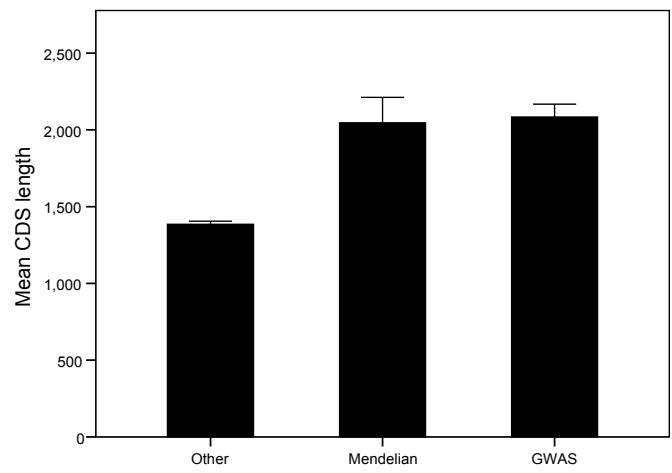

**Figure 1.** Mean CDS length. Error bars represent the 95% confidence intervals.

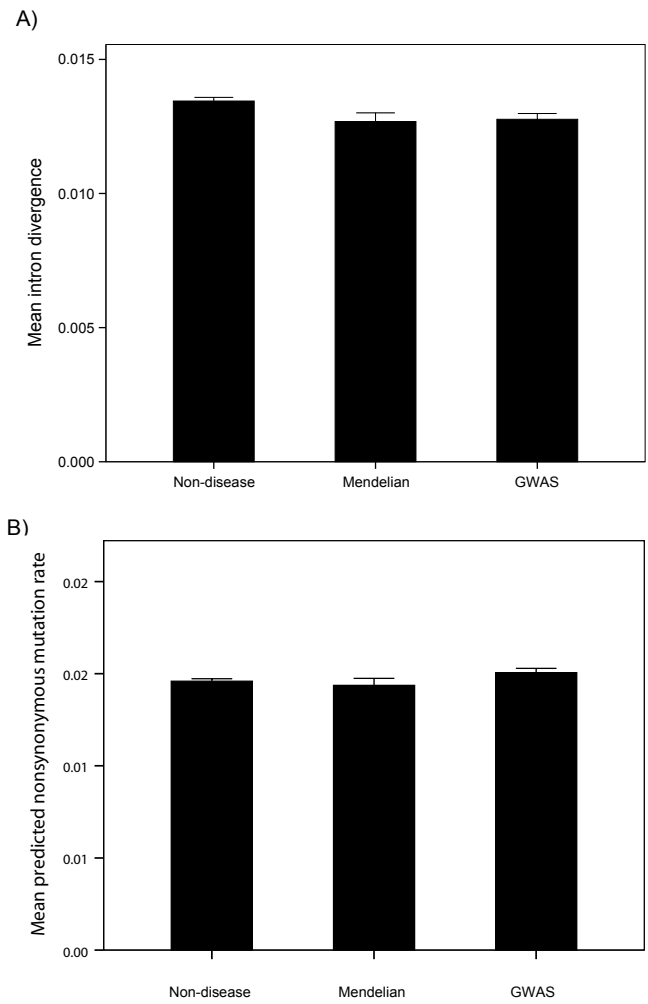

**Figure 2.** Mutation rates as inferred from intron divergence between human and chimpanzee. A) Intron divergence between human and chimpanzee; B) the predicted non-synonymous mutation rate. Error bars represent the 95% confidence intervals.

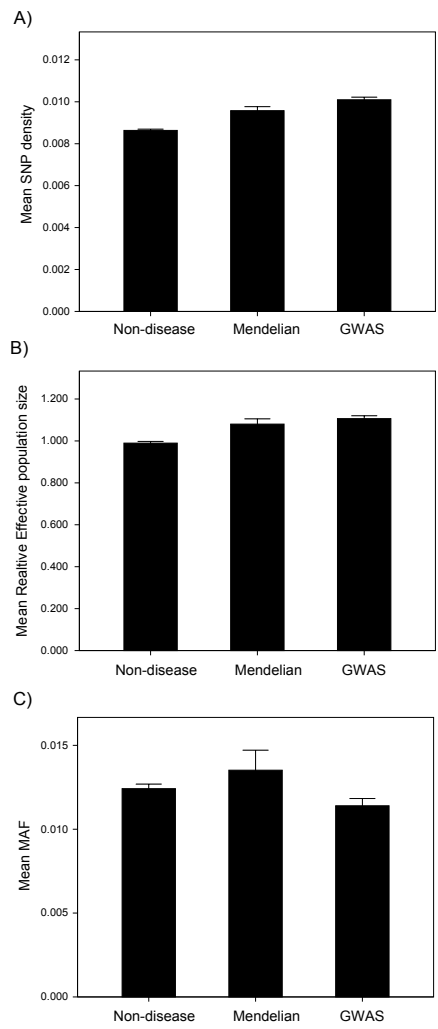

**Figure 3.** The diversity in disease and non-disease genes measured as the A) average intron SNP density, B) the average intron SNP density divided by intron divergence, C) and the mean minor allele frequency (MAF). Error bars represent the 95% confidence intervals.